\documentclass[12pt]{report}

\usepackage[dvips]{graphicx}
\usepackage{amssymb}
\usepackage{amsmath}

\makeatletter
\renewcommand{\@oddhead}{\hfil \thepage}
\renewcommand{\@oddfoot}{}
\renewcommand{\thefootnote}{\alph{footnote}}

\makeatother

\begin{document}

\renewcommand{\bibname}{\large \textit{References}}

\begin{center}
\textbf{\large {Confinement in a Model with Condensate of the
Scalar Field}}
\end{center}
\begin{center}
\textbf{V.E. Kuzmichev} \footnote{$\ $ e-mail:
specrada@bitp.kiev.ua} \textbf{and V.V. Kuzmichev} \footnote{$\ $
e-mail: specrada@bitp.kiev.ua; vvkuzmichev@yahoo.com}
\end{center}
\begin{center}
\textit{Bogolyubov Institute for Theoretical Physics, National
Academy of Sciences of Ukraine, Metrolohichna St. 14b, Kiev, 03143
Ukraine}
\end{center}
\setcounter{footnote}{0}
\renewcommand{\thefootnote}{\arabic{footnote}}

\textbf{Abstract:} The idea of ``soft'' confinement when the
lifetime of hadron with respect to quark-gluon channel of decay is
greater or at least of the order of some characteristic time for
our Universe is considered. Within the framework of a model of
three nonlinearly interacting fields the explicit form of an
effective potential is found. It provides the confinement of a
massive particle within the limited region of space by means of
constant component of the potential which arises as a result of
reorganization of vacuum of one of the scalar fields. It is shown
that the lifetime of hadron being equal to the age of the Universe
leads to the Higgs boson with the mass $m_{H} > 63.7$ GeV and
realistic self-constant.
\\[0.5cm]

\textbf{1. Introduction.} The experimental data and the
predictions of gauge theory of the strong interactions allow to
conclude that an effective coupling constant at sufficiently small
distances becomes small (phenomenon of asymptotic freedom) but at
distances large in comparison with the hadron size ($\sim 1$ fm)
the coupling becomes very strong and results in quark and gluon
confinement \cite{1}. In other words the channels of decay of
hadron via quarks and gluons are strictly forbidden. The nature of
this prohibition is still unknown and it is introduced into the
theory phenomenologically \cite{2}. In nonrelativistic
approximation a naive potential model gives a good description of
the radial excitations of "hidden charm" $(c\bar{c})$ mesons if at
large distances the potential is linearly or logarithmically
growing to infinity but at small distances it is similar to
Coulomb interaction \cite{1}. At the distances $\sim 1$ fm an
effective cut off of the confining potential must be realized. The
bag model \cite{3} - \cite{7} is an example of such model where
the condition that colour objects cannot appear as free particles
is introduced explicitly by means of disappearance of quark
current on the surface of the bag. With respect to potential model
such condition is equivalent to the introduction of the unnatural
infinite repulsion on the side of the external region to which
quarks cannot penetrate.

In the present article we show that above mentioned cut off can be
provided naturally by the condensate of the Higgs field which
arises from the reorganization of the vacuum state \cite{8} in the
system of nonlinearly interacting fields. In such approach from
the idea of absolute (strict) confinement (the decay of hadron
into the free quarks and gluons is strictly forbidden) we come to
the assumption of soft (nonstrict) confinement when the lifetime
of hadron with respect to quark-gluon channel of decay is greater
or at least of the order of some characteristic time for our
Universe. For example, the age of our Universe can be in the role
of such time.

\textbf{2. Theory.} We shall consider the system of three
nonlinearly interacting fields $\phi , \ \varphi_{1} \ \mbox{and}
\ \varphi_{2}$. The field $\phi $ is supposed to be massive and
complex and the fields $\varphi_{i}$ are real. For simplicity we
shall not take into account their spin, colour and flavour degrees
of freedom. The corresponding generalization can be made by
transition to spinor algebra. Let us define the Lorentz-invariant
Lagrangian in the form
\begin{equation}
\mathcal{L} = \partial_{\mu} \phi^{*}\,\partial^{\mu}\phi -
\phi^{*}m^{2}\phi +
\sum_{i=1}^{2}\left[\frac{1}{2}\,\partial_{\mu}\varphi_{i}\,\partial^{\mu}
\varphi_{i} - \phi^{*}U_{i}(\varphi_{i})\phi\right],
  \label{1}
\end{equation}
where $m$ is the mass of the field $\phi $. The functions
$U_{i}(\varphi_{i})$ are chosen in the form of anti-Higgs $(i =
1)$ and Higgs $(i = 2)$ potentials respectively
\begin{equation}
U_{i}(\varphi_{i}) = (-1)^{i}\left[-
\frac{\mu^{2}_{i}}{2}\,\varphi^{2}_{i} + \frac{\lambda_{i}}{4}\,
\varphi^{4}_{i}\right]
 \label{2}
\end{equation}
with constants $\mu^{2}_{i} > 0$ and $\lambda_{i} > 0$.
Variational principle being applied to the Lagrangian (\ref{1})
leads to the set of three nonlinear equations:
\begin{equation}
\square \phi = - \left[\sum_{i} U_{i} + m^{2} \right]\phi,
  \label{3}
\end{equation}
\begin{equation}
\square \varphi_{i} = - \left|\phi \right|^{2}\,\frac{\partial
U_{i}}{\partial \varphi_{i}}.
  \label{4}
\end{equation}
We shall limit ourselves to the consideration of the excitation
spectrum of the field $\phi$ with energies $E' \ll 2 m$. After the
unitary transformation $\phi = \psi \exp(-i m t)$ the equation
(\ref{3}) in mentioned nonrelativistic limit reduces to the
Schr\"{o}dinger equation
\begin{equation}
i \, \partial_{t} \, \psi = \frac{1}{2 m}\left[-\nabla^{2} +
V\right] \psi
  \label{5}
\end{equation}
with the potential
\begin{equation}
V = \sum_{i} U_{i}(\varphi_{i}).
  \label{6}
\end{equation}
This equation describes the motion of a particle with the mass $m$
in the field of the potential (\ref{6}) being provided by the
relativistic fields $\varphi_{i}$. We shall consider the case when
a particle scatters on the potential (\ref{6}) or tunnels from it.
We take the axis $OZ$ for the initial direction of its motion. It
turned out that the interaction will have the form of the barrier
interesting for us if we suppose that dissipation of the fields
$\varphi_{i}$ in $(x, y)$ - plane is small. It means that the
fields $\varphi_{i}$ represent the structures stretched along the
axis $OZ$ with a cross-section $R^{2} = x^{2} + y^{2}$ small in
comparison with the square of the distance $z^{2}$ at which the
physical processes under consideration are observed. In this case
the radial variable $r$ in the equation (\ref{5}) coincides with
$|z|$ to within the terms of the order of $R^{2} / 2 \, z^{2}$ and
fields $\varphi_{i}$ form the local spacetime particle-like
formations which are the carriers of the effective interaction
along the $r$. Considering the amplitude $|\phi| = |\psi|$ as a
slowly varying function of the variables $(x, y, z ; t)$ in the
above approximation we shall find the solutions of the equations
(\ref{4}) in the form of the solitary wave of $\mbox{sech}$-type
$(i = 1)$ and kink/antikink $(i = 2)$ in $(z, t)$ - plane \cite{9,
10}. Substitution of these solutions into (\ref{6}) leads to the
effective potential
\begin{equation}
V(z,t) = \frac{\mu^{4}_{1}}{\lambda_{1}}\, \mbox{sech}
^{2}\left(\mu_{1}|\psi|s_{1}\right) \tanh ^{2}
(\mu_{1}|\psi|s_{1}) + \frac{\mu ^{4}_{2}}{4
\lambda_{2}}\left[\mbox{sech}^{4}\left(\frac{\mu_{2}}{\sqrt{2}}|\psi|s_{2}\right)
- 1 \right],
  \label{7}
\end{equation}
where $s_{i} = (z - u_{i} t) / \sqrt{1 - u_{i}^{2}}$, and $u_{i}$
are the velocities of the solitary waves, $-1 < u_{i} < 1$.

The first term in (\ref{7}) is proportional to the sech-wave
energy density being local. It has the shape of two humps with the
maximums in the points $\mu_{1} | \psi | s_{1} = \pm 0.881$ and
minimums at $\pm \infty$ and at the origin and organizes the
interaction in the form of the barrier of the finite magnitude.

The second term in (\ref{7}) describes the part of the interaction
formed by the kink/antikink with the energy density localized near
the value $|\psi| \, s_{2} = 0$. Within limit of  $|\psi| \, s_{2}
\rightarrow \pm \infty$ it takes the constant negative value
$V_{c} = - \mu_{2}^{4} / 4 \, \lambda_{2}$.

Since both terms in (\ref{7}) in the $(z, t)$ - plane have a
pronounced local character then the possible dependence of the
amplitude $|\psi|$ on these variables does not change general form
of the potential $V(z, t)$. Its concrete form depends only on the
relations between the parameters $\mu_{i},\ \lambda_{i}\
\mbox{and} \ u_{i}$. In stationary case at $(\mu_{1}^{4} / \lambda
_{1}) > (\mu_{2}^{4} / \lambda _{2})$ it has the form of the
barrier localized in the space. One of the possible variants is
shown in Fig. 1.

\begin{figure}
\begin{center}
\includegraphics[scale=1]{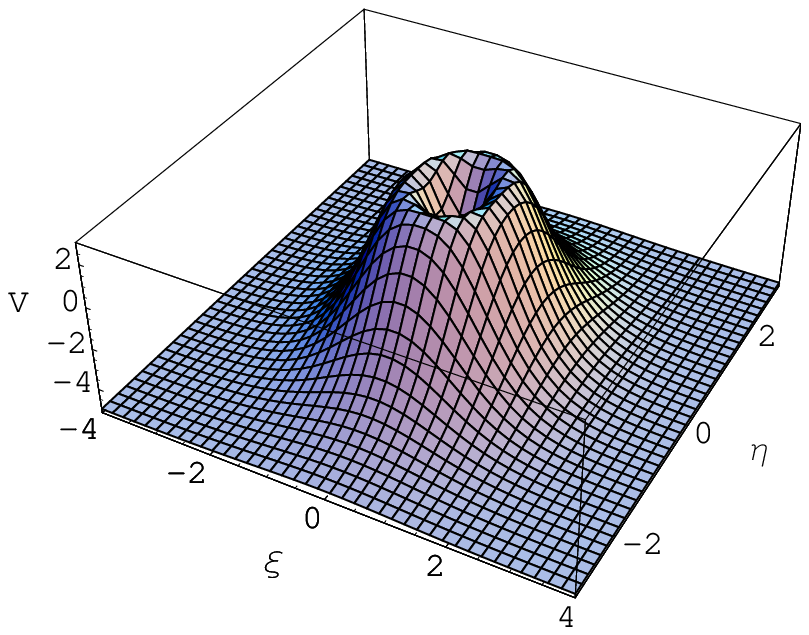}
\includegraphics[scale=1]{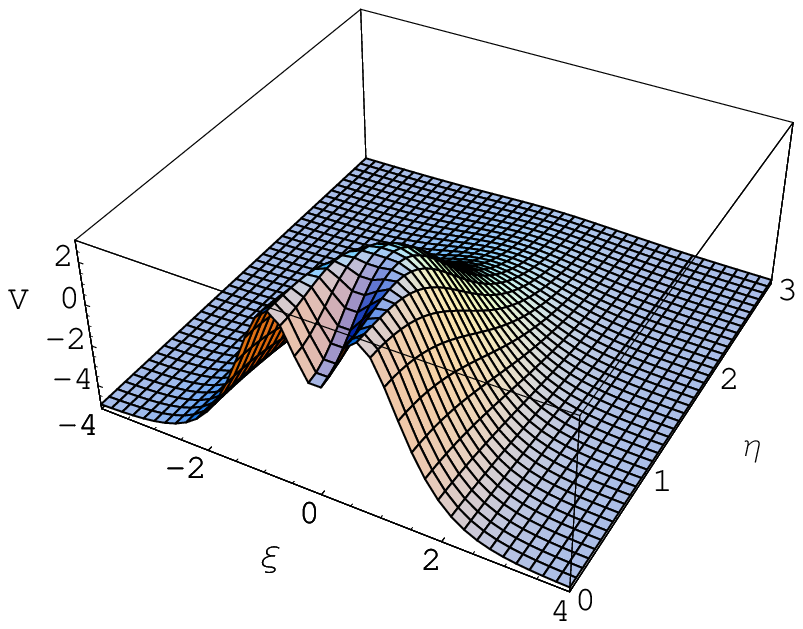}
\end{center}
\vspace{-2.cm} \caption{The effective potential (\ref{7}) in the
stationary case ($u_{i} = 0$) as a function of the radial variable
$r = \sqrt{\xi^{2} + \eta^{2}}$ and its section by the plane
passing across the axis of symmetry.}
\end{figure}

Scattering of the particle on such potential is described by the
wave function in the form of the superposition of converging and
diverging waves. It is convenient to normalize it to the unit
density of the incident flux, $\psi = \exp (i q z - i E t) +
\psi_{scat}$, where $E = q^{2}/2 m$ is the energy of the particle
and $\psi_{scat}$ takes into account the action of the potential.
Since $\psi _{scat} \rightarrow 0$ at $r \rightarrow 0$ \cite{11}
then $|\psi| \rightarrow 1$ at $r \rightarrow \infty$. Hence at
large values of $|z|$ the potential (\ref{7}) tends to the
constant negative value $V_{c}$ which can be expressed in terms of
the Higgs mass $m_{H} = 2 \, \mu_{2} \, m$ and Higgs self-constant
$\lambda = \lambda_{2} \, m^{2}$ corresponding to the free field
$\varphi_{2}$,
\begin{equation}
V_{c} = - \frac{1}{64 \, \lambda} \, \frac{m_{H}^{4}}{m^{2}}.
  \label{8}
\end{equation}
The appearance of the constant component (\ref{8}) of the
potential (\ref{7}) at large distances $|z|$ can be interpreted as
a consequence of the reorganization of vacuum of the field
$\varphi_{2}$ as a result of spontaneous symmetry breaking. Its
value is determined by the constant field $v$ (condensate)
originating from such reorganization and being equal to
\begin{equation}
v = \pm \, \frac{m_{H}}{2 \, \sqrt{\lambda}} = \pm \,
\frac{\overline{m}_{H}}{2 \, \sqrt{\overline{\lambda}}},
  \label{9}
\end{equation}
where $\overline{m}_{H} = m_{H}\,|\psi|$ and $\overline{\lambda} =
\lambda \,|\psi|^{2}$ are the effective values of the Higgs mass
and the Higgs self-constant in the presence of the field $\psi$
(which generally speaking depend on the energy $E$ and on the
location in the spacetime).

\textbf{3. Numerical example.} In order to demonstrate the role of
the potential $V_{c}$ in the process of quantum tunneling of the
particle from the region inside the barrier we shall consider the
stationary case ($u_{i} = 0$) and replace corresponding ``real''
effective potential by the simpler one in the form of the sum of
square steps \cite{12}
\begin{equation}
V = V_{1}\,\theta (r - r_{1})\,\theta (r_{2} - r) -
V_{2}\,\theta(r - r_{2}),
  \label{10}
\end{equation}
where $V_{1}$ and $V_{2}$ are the positive constants, $r_{2}$ is
the size of the barrier, $\Delta r = r_{2} - r_{1}$ is its width
and $\theta (x)$ is the step-function.

Since we are interested only in the integral characteristics of
the process then we may impose the constraints on the parameters
of the potentials (\ref{7}) and (\ref{10})
\begin{equation}
V_{1} = \frac{1}{\Delta r} \, \int_{0}^{r_{0}}\!\!dz
\overline{V(z, t)}, \quad \quad V_{2} = - V_{c},
  \label{11}
\end{equation}
where $\overline{V(r_{0}, t)} = 0$ and the bar denotes transition
to the stationary case. The quantities $\Delta r$ and $r_{1}$ are
determined by the corresponding parameters of the hadron. Such
simple model, of course, does not describe observed spectroscopy
of mesons but it has two important properties: (i) there exists
the domain $r < r_{1}$ where the quarks with the energy $E <
V_{1}$ behave as free; (ii) at definite relation between $V_{1}$
and $V_{2}$ the quark tunneling into the range $r > r_{2}$ is
suppressed. Indeed the direct calculations of the decay
probability of the system from $S$ - state with the energy close
to the energy of the stationary state $E_{0} = k_{0}^{2} / 2 m$
determined by the transcendent equation
\begin{equation}
\sqrt{V_{1} - k_{0}^{2}}\,\tan k_{0}r_{1} = - k_{0}
  \label{12}
\end{equation}
leads to the width for decay into quarks
\begin{equation}
\Gamma = \frac{8}{m}\,\frac{k_{0}^{2}\,(V_{1} -
k_{0}^{2})^{3/2}\,(V_{2} + k_{0}^{2})^{1/2}}{V_{1}\,(V_{1} +
V_{2})\,(1 + \sqrt{V_{1} - k_{0}^{2}}\,r_{1})} \, \exp
\left\{-\,2\,\sqrt{V_{1} - k_{0}^{2}}\,\Delta r \right\}.
  \label{13}
\end{equation}
At $V_{1} \sim V_{2}$ the width $\Gamma$ is determined mainly by
the exponential factor. But at $V_{2} \gg V_{1}$ the
pre-exponential multiplier becomes essential. The width $\Gamma$
decreases as $V_{2}^{-1/2}$ when $V_{2}$ grows and for
sufficiently large value of $V_{2}$ but finite nonzero exponential
factor we have $\Gamma \sim 0$.

Let us estimate $V_{1}$ and $V_{2}$ for some ``realistic'' values
of $\Gamma$ in approximation $V_{1} \gg k_{0n}^{2} =
\pi^{2}\,n^{2} / r_{1}^{2}$, where $n = 1, 2, 3 \ldots $ is the
number of the $S$ - level. We shall consider the tunneling of the
quark from the ground $1S$ - state. In order to illustrate the
role of $V_{2}$ at first we assume that $V_{2} = 0$ and the width
of the $1S$ - level equals to the decay constant of the nucleus
$\mbox{U}_{92}^{238}$ into $\alpha$ - particle and
$\mbox{Th}_{90}^{234}$, $\Gamma_{\alpha} = 7 \times
10^{-18}\,\mbox{s}^{-1}$. Taking the quark mass, the hadron size
and the range of barrier to be equal to $m = 350$ MeV, $r_{1} = 1$
fm and $\Delta r = 0.4$ fm respectively, we obtain $V_{1} = 7
\times 10^{2}$ GeV for $\Gamma = \Gamma_{\alpha}$ and the energy
$E_{01} = 547$ MeV for emitted quark. Since free quarks are not
observed then the lifetime of hadron with respect to quark decay
mode is not less then the age of the Universe with $\Gamma_{0} = 2
\times 10^{-18}\,\mbox{s}^{-1}$. For obtained $V_{1}$ the value
$\Gamma = \Gamma_{0}$ is reached at $V_{2} \approx 10^{7}$ GeV.
The condition $E_{01} \ll V_{1} \ll V_{2}$ is valid. If hadron
decays with respect to the same quark channel during the time
typical for weak interactions then for $\Gamma_{w} =
10^{8}\,\mbox{s}^{-1}$ and $V_{2} = 0$ we would have $V_{1} = 95$
GeV. At such height of barrier $V_{1}$ the width $\Gamma = \Gamma
_{0}$ is reached at $V_{2} \approx 10^{55}$ GeV.

These estimations show that along with practically infinite values
(such as $V_{1} = \infty, \ V_{2} \sim 0$, corresponding to quark
bag model, or $V_{1} < \infty, \ V_{2} = \infty$) there also exist
the values of $V_{1}$ and $V_{2}$, which provide the quark
confinement while remaining finite within the energy scale of high
energy physics.

Knowing $V_{2}$ we can estimate the value of admissible Higgs
mass. For the value $V_{2} \approx 10^{7}$ GeV and quark mass $m =
350 $ MeV we have the following relation between the values
$m_{H}$ and $\lambda$,
\begin{equation}
m_{H} \approx 272 \, \lambda ^{1/4} \ \mbox{GeV}.
  \label{14}
\end{equation}
For $\lambda > 3 \times 10^{-3}$ \cite{13} it gives the value
$m_{H} > 63.7$ GeV which coincides with the current theoretical
and experimental estimations of the Higgs mass \cite{14}.\\[0.5cm]

We should like to express our gratitude to Alexander von Humboldt
Foundation (Bonn, Germany) for the assistance during the research.

\end{document}